\documentclass[10pt, conference, compsocconf]{IEEEtran}
\usepackage{cite}
\usepackage{amsmath,amssymb,amsfonts}
\usepackage{algorithmic}
\usepackage{graphicx}
\usepackage{textcomp}
\usepackage{xcolor}
\usepackage[bookmarks=false]{hyperref}
\usepackage{booktabs} % For formal tables
\usepackage[lined,boxruled,norelsize,linesnumbered]{algorithm2e}
\usepackage{soul}
\def\BibTeX{{\rm B\kern-.05em{\sc i\kern-.025em b}\kern-.08em
    T\kern-.1667em\lower.7ex\hbox{E}\kern-.125emX}}

\usepackage[para]{footmisc}
\usepackage{dblfloatfix}
\usepackage{bigdelim}

\usepackage[utf8]{inputenc}
\usepackage[T1]{fontenc}
\usepackage{microtype}
\usepackage{xspace}
\usepackage{multirow}

\usepackage{dblfloatfix}    % To enable figures at the bottom of page

\usepackage{flushend}

\usepackage{footmisc}

\usepackage{tikz}

\usepackage{graphicx,color}
\usepackage{soul}
	
\usepackage{color, colortbl}

\newcommand{\reportOnly}[1]{}

\newcommand{\REMOVE}[1]{}

\newcommand{\NO}[1]{}

\usepackage{url}

\usepackage{tikz,xcolor,hyperref}

\definecolor{lime}{HTML}{A6CE39}
\DeclareRobustCommand{\orcidicon}{
	\begin{tikzpicture}
	\draw[lime, fill=lime] (0,0) 
	circle [radius=0.16] 
	node[white] {{\fontfamily{qag}\selectfont \tiny ID}};
	\draw[white, fill=white] (-0.0625,0.095) 
	circle [radius=0.007];
	\end{tikzpicture}
	\hspace{-2mm}
}
\foreach \x in {A, ..., Z}{\expandafter\xdef\csname orcid\x\endcsname{\noexpand\href{https://orcid.org/\csname orcidauthor\x\endcsname}
			{\noexpand\orcidicon}}
}

\begin{document}

%\pagenumbering{arabic} 
%\pagestyle{plain}
%\bstctlcite{IEEEexample:BSTcontrol}

\title{\plugin: A Web Gamification Tool to Improve Locators and Page Objects Quality}
%\titlerunning{BEWT: A Benchmark for End-to-End Web Testing}

\newcommand{\orcidauthorA}{0000-0002-6618-4186}
\newcommand{\orcidauthorB}{0000-0001-5267-0602}
\newcommand{\orcidauthorC}{0000-0002-3928-5408}
\newcommand{\orcidauthorD}{0000-0001-7651-0400}

\newcommand{\plugin}{\textsc{TestQuest}\xspace}

\author{
%\IEEEauthorblockN{Dario Olianas\orcidA{}, Diego Clerissi\orcidD{}, Maurizio Leotta\orcidB{}, Filippo Ricca\orcidC{}}
\IEEEauthorblockN{Dario Olianas, Diego Clerissi, Maurizio Leotta, Filippo Ricca}
\IEEEauthorblockA{Dipartimento di Informatica, Bioingegneria, Robotica e Ingegneria dei Sistemi (DIBRIS), Universit\`a di Genova, Italy\\
dario.olianas@dibris.unige.it, diego.clerissi@dibris.unige.it, maurizio.leotta@unige.it, filippo.ricca@unige.it}
}

 \maketitle

\begin{abstract}
Web applications play a crucial role in our daily lives, making it essential to employ testing methods that ensure their quality. Typically, Web testing automation frameworks rely on locators to interact with the graphical user interface, acting as connection points to the elements on a Web page. Nevertheless, locators are widely recognized as a major vulnerability in Web testing, as they are highly sensitive to the frequent changes in Web page structures caused by rapid software evolution. 
The adoption of the Page Object pattern to separate test logic from structural layout -- supporting code reuse and maintainability -- has generally led to more robust test cases. However, their implementation is a manually intensive task, and even automated support may require manual realignment efforts.

Although gamification strategies have recently been integrated into the Web testing process to boost user engagement, using tasks and rewards aligned with testing activities, they have not yet been employed to enhance the robustness of locators and support the implementation of Page Objects. In this paper, we introduce \plugin, a tool designed to improve test robustness by applying gamification to locators and Page Objects, boosting user engagement while guiding them toward the adoption of best practices. %\ML{aggiungere Po}
\end{abstract}


\section{Introduction}\label{sec:introduction}The pervasive presence of Web applications in our daily lives demands effective testing techniques to ensure that test artifacts remain aligned with the fast pace of software evolution~\cite{leotta2013comparing}. Among the most commonly used Web testing tools~\cite{2023-leotta-ICST} there is Selenium~\cite{selenium}, which offers APIs for interacting with Web elements through locators -- references that identify elements using various strategies applied to the Document Object Model (DOM) of Web pages (e.g., by name or XPath descriptor).

Manually writing test cases, even with the support of test-assistance tools, can be a time-consuming and demanding task. To address this, automated test generation techniques can be leveraged -- ranging from simple random exploration to more advanced methods, such as systematic Web page exploration, scriptless approaches based on GUI changes, and reinforcement learning algorithms~\cite{clerissi2024guess}. However, most of these techniques depend on locators, which are considered one of the primary sources of fragility in Web testing~\cite{hammoudi2016record}, as they are tightly coupled with the structural aspects of the DOM. As a result, even minor GUI changes during software evolution can disrupt the connection between locators and the corresponding Web elements. Maintaining locator robustness during software evolution -- and repairing them when they break -- demands considerable effort in identifying the most robust locator strategies, making the process both time-consuming and tedious. Although there are techniques for automatically generating locators using robustness heuristics, they often still require iterative manual adjustments to align with the specific application under test, for instance to support parametric test scripts.~\cite{leotta2016robula+,leotta2021sidereal,nguyen2021generating}.

To aid in Web testing, the Page Object design pattern~\cite{pageobject,leotta2020family} was introduced as an effective approach that adds a layer of indirection between the test logic and the structure of Web pages. This allows test scripts to remain independent of structural details, which are instead encapsulated within page-specific abstractions. This intermediate layer helps ensure that changes to Web pages demand minimal modifications to the test code, since element locators are centralized within Page Objects, promoting the creation of more robust and maintainable test artifacts. Nonetheless, manual testing efforts persist, as Page Objects may still require ad-hoc fixes and design adjustments.

The gamification paradigm has emerged as a way to incorporate game-like elements into the testing domain, with the aim of making testing activities more engaging~\cite{de2018gamification}. By aligning tasks and rewards with established patterns and best practices, gamification can encourage more frequent testing and stimulate testers' creativity in crafting high-quality test artifacts. It also holds potential for educational use, helping to convey key testing concepts~\cite{blanco2023can}. However, most of the research has concentrated on unit testing, with comparatively little attention given to Web and GUI testing~\cite{fulcini2023review,de2018gamification}.

%To address this, we present \plugin, a plugin for IntelliJ IDEA designed to enhance locators and Page Objects quality in Web testing, originally sketched in our previous work~\cite{clerissi2024gaming}. 
For this reason, we decided to focus on the development of \plugin, a plugin for IntelliJ IDEA designed to enhance the quality of locators and Page Objects in web testing, originally outlined in our previous work\cite{clerissi2024gaming}.

By embedding a gamification framework directly into the testing environment, \plugin motivates users to develop more resilient locators through incentives such as rewards, achievements, and customizable profiles. To the best of our knowledge, gamification has not yet been applied in the context of locators robustness, nor in the context of the implementation of the Page Object pattern. %\ML{anche qua menzione la parte sui PO} 
\section{Background}\label{sec:background}%\subsection{Gamification}

Gamification is an emerging paradigm in both academia and industry, aiming to incorporate game elements into non-game contexts to improve user engagement and make repetitive or tedious tasks more enjoyable~\cite{deterding2011gamification}.

%For example, during requirements elicitation, it can motivate stakeholders through voting systems that aid in clarifying requirements or by rewarding those who identify new requirements. In software development, developers might earn rewards based on metrics from code quality tools like SonarQube~\footnote{https://www.sonarsource.com/products/sonarqube}, or according to their tangible contributions to the team. 

At its core, gamification seeks to deliver an engaging and interactive experience by encouraging users to complete tasks that offer a sense of challenge and fulfillment, delivering rewards to reflect their progress. 

%A central element of this approach is the customizable user profile, which allows individuals to personalize their identity and showcase rewards earned through task completion.

%Rewards can include experience points that unlock new levels and titles, reflecting the user's skills and status. Additionally, users may earn icons that serve as visual trophies, which can be displayed on their profiles. 

Users can be assigned various tasks, such as \textit{achievements} and \textit{dailies}. Achievements serve as a core driver of the gamification experience, encouraging users to pursue challenging tasks, whereas dailies are more approachable and often time-limited tasks that introduce a level of unpredictability and help maintain user engagement on a daily basis.

%Each achievement usually includes a distinctive name, a clear description of the goal, an illustrative icon, and a reward granted upon completion. Achievements often involve completing challenging or time-consuming tasks, rewarding users with experience points and exclusive icons upon success. Dailies, instead, are time-limited achievements randomly assigned to users. Generally less demanding, they introduce an element of unpredictability and help maintain user engagement even after all standard achievements have been completed.

Gamification has also gained momentum in software engineering, as it can enhance various stages of the development process. 
However, according to the literature, gamification applied to testing has mostly neglected the Web domain~\cite{de2018gamification,fulcini2023review}. 

In Web testing, a locator identifies a Web element within a page through a strategy (e.g., XPath)\footnote{https://www.selenium.dev/documentation/webdriver/elements/locators/} and a corresponding value (e.g., \texttt{\small /html/body/div[3]/div/form/div[1]/input}). In the literature~\cite{leotta2013comparing}, locator failures remain a leading cause of test fragility~\cite{hammoudi2016record}, often requiring manual and costly repairs.
Several approaches have been proposed to automatically generate or repair locators using heuristics and statistical techniques, aimed at producing resilient locators~\cite{leotta2021sidereal,kirinuki2019color,fasolino2022towards}.
Furthermore, the Page Object pattern has also been widely used as a solution to improve test maintainability and code reuse~\cite{pageobject,leotta2020family}, by isolating the references to locators and page structure in object-oriented classes, receiving support from automated approaches for code generation~\cite{yu2015incremental,stocco2017apogen,leotta2022assessor} through Web crawling or test recording. 

As all these methods involve high computational processing, or require manual adjustment for readability and stability, gamification may naturally alleviate the costs of their application. 

A framework to employ gamification to GUI testing was proposed and validated by Coppola \textit{et al.}~\cite{coppola2023effectiveness}, defining game mechanics to evaluate the tester performance. 
Another proposal is GERRY~\cite{garaccione2022gerry}, a Google Chrome extension that enhances Capture \& Replay GUI testing with gamification elements and reporting. An IntelliJ plugin named GIPGUT~\cite{Garaccione2024GamifiedGT} was proposed to influence test effectiveness, by designing tasks for Selenium WebDriver test implementation and execution. 

At present, no existing techniques explicitly adapt gamification to Locators and Page Objects engineering, even though these components are central to Web test quality. 
\section{\plugin plugin}\label{sec:design}

\begin{figure}[h!]
\vspace{2mm}
 \centering
\includegraphics[width=0.99\linewidth, clip, trim=0 110 70 0]{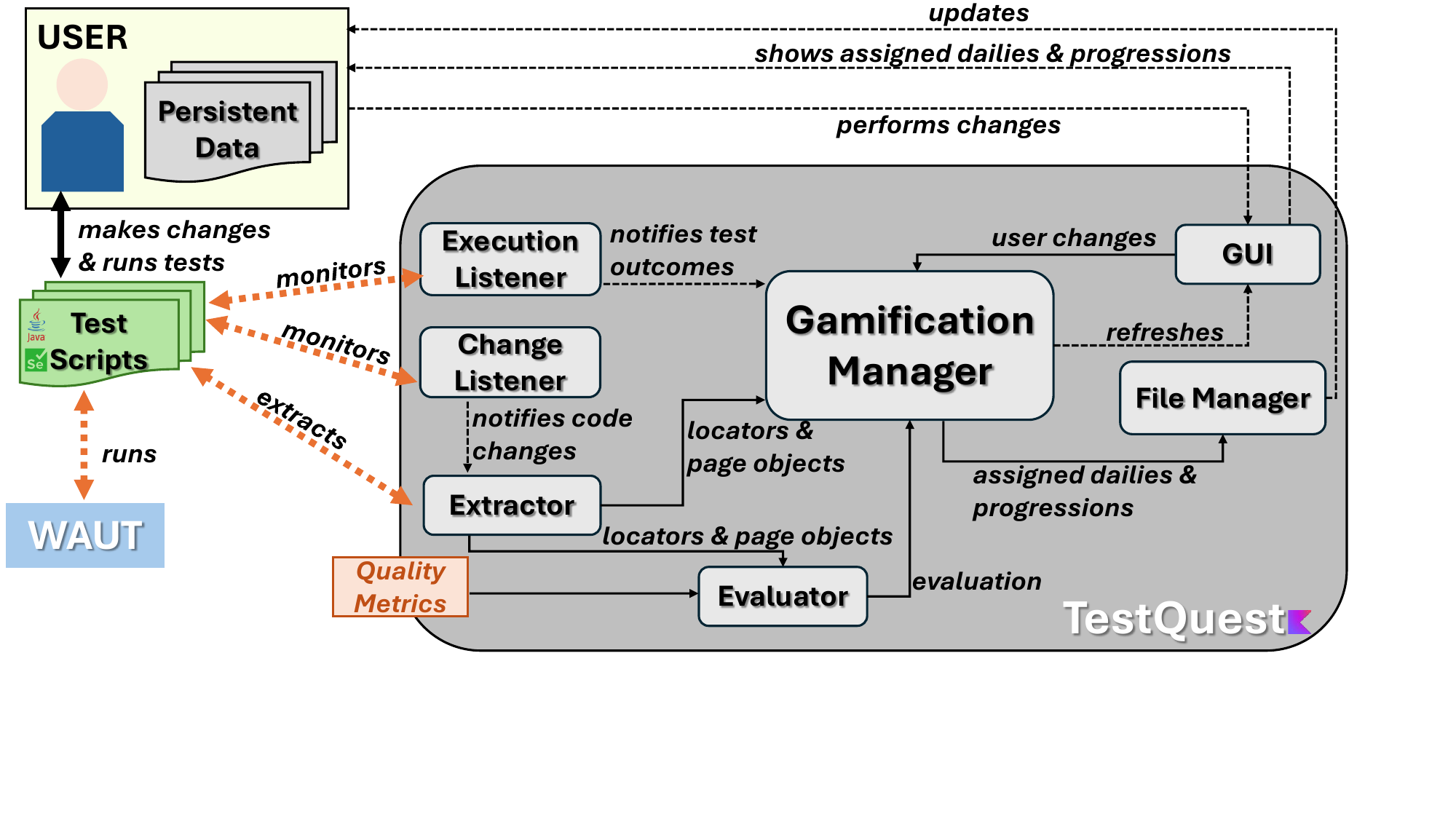}
\vspace{-5mm}
     \caption{The \plugin workflow and architecture.}
     \label{fig:testquest}
     \vspace{-3mm}
\end{figure}

\plugin is a plugin for IntelliJ IDEA, developed in Kotlin, designed to enhance test robustness through an embedded gamification framework. \plugin plugin, with installation and usage instructions, can be found at \texttt{\small\url{https://gitlab.com/kandriann/testquest}}.
%
% The workflow of \plugin is described in Figure~\ref{fig:approach}
%
% \begin{figure}[h!]
%     \includegraphics[width=10cm, clip, trim=30 50 30 30]{img/TestQuest.pdf}
%     \caption{The \plugin approach.}
%     \label{fig:approach}
% \end{figure}
%
\plugin relies on a task-driver approach, where tasks to complete are based on test robustness best practices identified in the literature, which we synthesized in Table \ref{tab:guidelines}, targeting Locators (\textbf{L1}-\textbf{L6})~\cite{leotta2013comparing,hammoudi2016record,kirinuki2019color,leotta2016robula+,leotta2021sidereal,di2024towards,nguyen2021generating,fulcini2022guidelines} and Page Objects (\textbf{P1}-\textbf{P6})~\cite{stocco2017apogen,selenium-pageobject,leotta2013comparing,pageobject,van2015testing}.
\begin{table}[t!]
\centering
\vspace{-2mm}
\caption{Literature-Based Best Practices for Test Robustness.}
\vspace{-1mm}
\label{tab:guidelines}
%\resizebox{\columnwidth}{!}{
\begin{sf}\footnotesize
%\tiny
{\def\arraystretch{1.0}
\setlength{\tabcolsep}{1.0pt}
%\begin{tabular}{r@{\,}c|l}
\begin{tabular}{r@{\,}c|>{\raggedright\arraybackslash}p{7.75cm}}
\cline{2-3}
& \textbf{ID} & \textbf{Description} \\
\cline{2-3}
{\multirow{9}{*}{\rotatebox[origin=c]{90}{Locator}}}\ldelim\{{9}{*}\parbox[t]{3mm}
& \textbf{L1} & Prioritize \textit{ID} and \textit{XPath} locators\\
& \textbf{L2} & Prioritize \textit{XPath} locators with predicates about \textit{id}, \textit{name}, \textit{class}, \textit{title}, \textit{alt}, and \textit{value} properties \\
& \textbf{L3} & Keep number of positional predicates and levels in XPath locators as few as possible \\
& \textbf{L4} & Keep locator values readable and short \\
& \textbf{L5} & Avoid using absolute \textit{XPath} locators \\
& \textbf{L6} & Avoid \textit{XPath} locators with predicates about internal app structure (e.g., \textit{href}) or Javascript code (e.g., \textit{onClick}) \\
\cline{2-3}
{\multirow{10}{*}{\rotatebox[origin=c]{90}{PageObject}}}\ldelim\{{10}{*}\parbox[t]{3mm}
& \textbf{P1} & Avoid exposing locator details outside Page Objects \\
& \textbf{P2} & Avoid implementing unused locators within Page Objects\\
& \textbf{P3} & Avoid methods implementing test logics in Page Objects (e.g., assertions, conditional statements) \\
& \textbf{P4} & Implement multiple Page Object methods to model multiple expected outcomes (e.g., \textit{loginOK} and \textit{loginKO}) \\
& \textbf{P5} & Introduce Page Object ancestors to share common functionalities \\
& \textbf{P6} & Add Page Objects as return types to methods to model the user's exploration \\
\cline{2-3}
\end{tabular}
\vspace{-5mm}
}\end{sf}
%}
\end{table}
%
%
% Once the tasks are assigned, the tester can then try to earn the proposed achievements and dailies by performing them. \plugin operates from two complementary perspectives: a \textit{static} perspective, which monitors changes in test scripts to track the tester’s progress toward task completion; and a \textit{dynamic} perspective, which analyzes test execution outcomes to validate those changes and update the tester’s profile accordingly. \plugin focuses on changes in test artifacts related to locators and Page Objects. These are periodically extracted from the test artifacts and compared across versions before and after the tester’s modifications.
%
% \plugin is currently implemented in Kotlin, with the persistence layer saving all users data and progressions managed via XML files, while the target test artifacts are expected to be developed in Java, using Selenium WebDriver~\cite{selenium} and JUnit libraries. 
% The plugin, with installation and usage instructions, can be found at \url{https://gitlab.com/kandriann/testquest}.
%
%\subsection{\plugin Architecture}
%
A conceptualization of \plugin workflow and its internal architecture is sketched in Figure~\ref{fig:testquest}.
\plugin main class is implemented as an IntelliJ custom action, which adds the option to activate the plugin in the target test project IDE\footnote{https://plugins.jetbrains.com/docs/intellij/action-system.html}.
Initially, any test-related file within the target project is retrieved, relying on extractors to extract data about locators and Page Objects. From a programmatic perspective, in \plugin the concept of locator is strongly inspired by the structure supported by Selenium WebDriver~\cite{selenium}, while the structure of a Page Object follows the design principles found in the literature~\cite{pageobject,selenium-pageobject}. Thus, locators are modeled as Kotlin data classes containing information such as \textit{locator type} (e.g., XPath), \textit{locator value} (e.g., "submitBtn") and \textit{placement} in the test suite (i.e., class, method, and line of code). Similarly, Page Objects are characterized by \textit{name}, \textit{ancestors}, and \textit{method list}, where each method is described by \textit{name}, \textit{return type}, \textit{associated locators}, and more. 

The extracted data are used to evaluate the test suite according to the quality metrics listed in Table~\ref{tab:guidelines}, and to instrument the gamification environment, establishing the tasks to assign to the user. 
The current implementation of \plugin supports a total of 50 dailies (30 about locators, 20 about Page Objects) and 29 achievements to motivate users by making progress, rewarding positive behavior, and encouraging them to explore less obvious aspects of locators and Page Objects mechanics.

\section{\plugin usage}\label{sec:usage}

\begin{figure}[t!]
\vspace{-1.5mm}
    \includegraphics[width=\columnwidth, clip, trim=0 0 260 0]{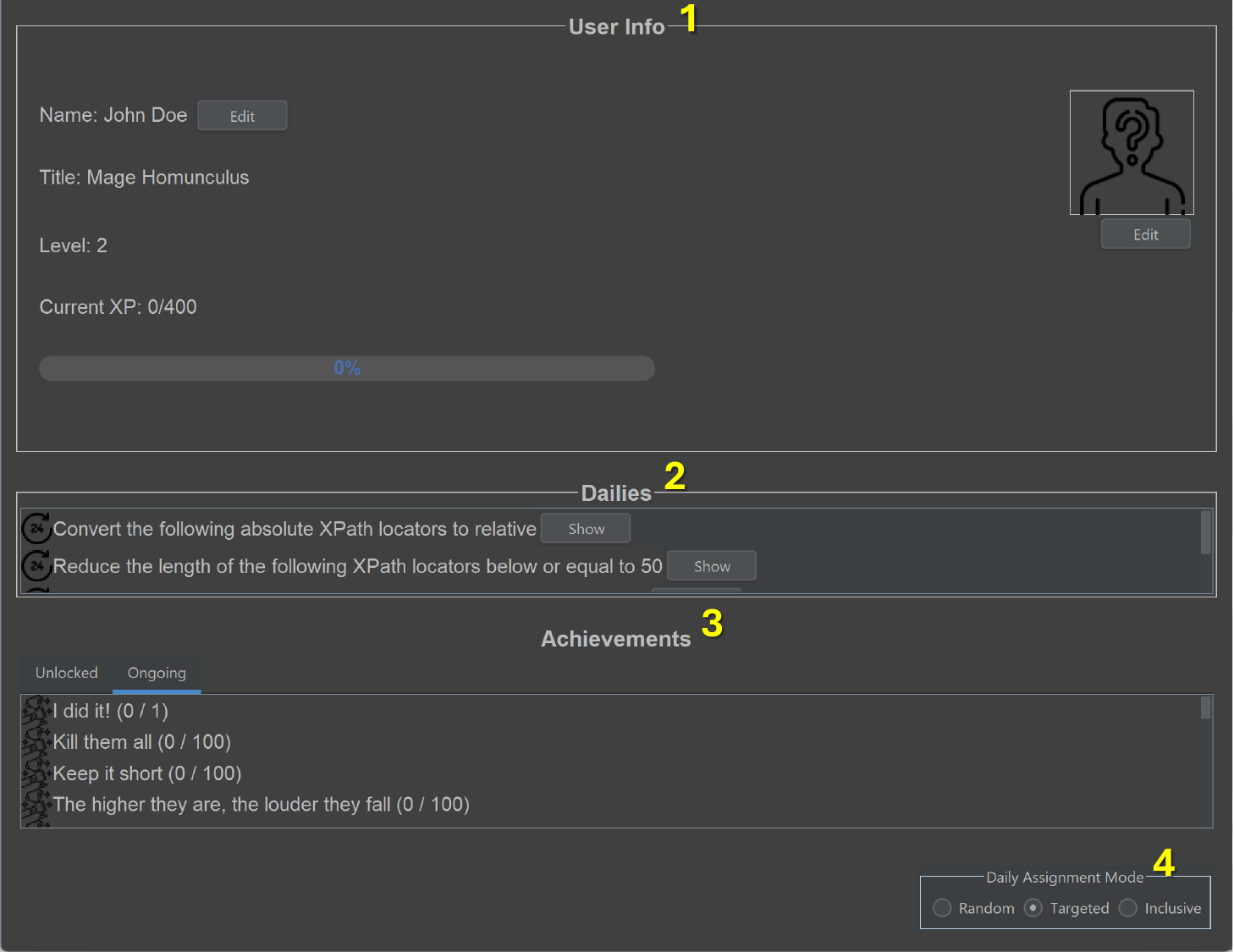}
    \vspace{-7mm}
    \caption{\plugin Gamification Window.}
    \label{fig:gamification-window}
    \vspace{-2mm}
\end{figure}

\begin{figure}[t!]
\vspace{-2mm}
    \includegraphics[width=\columnwidth, clip, trim=0 95 0 0]{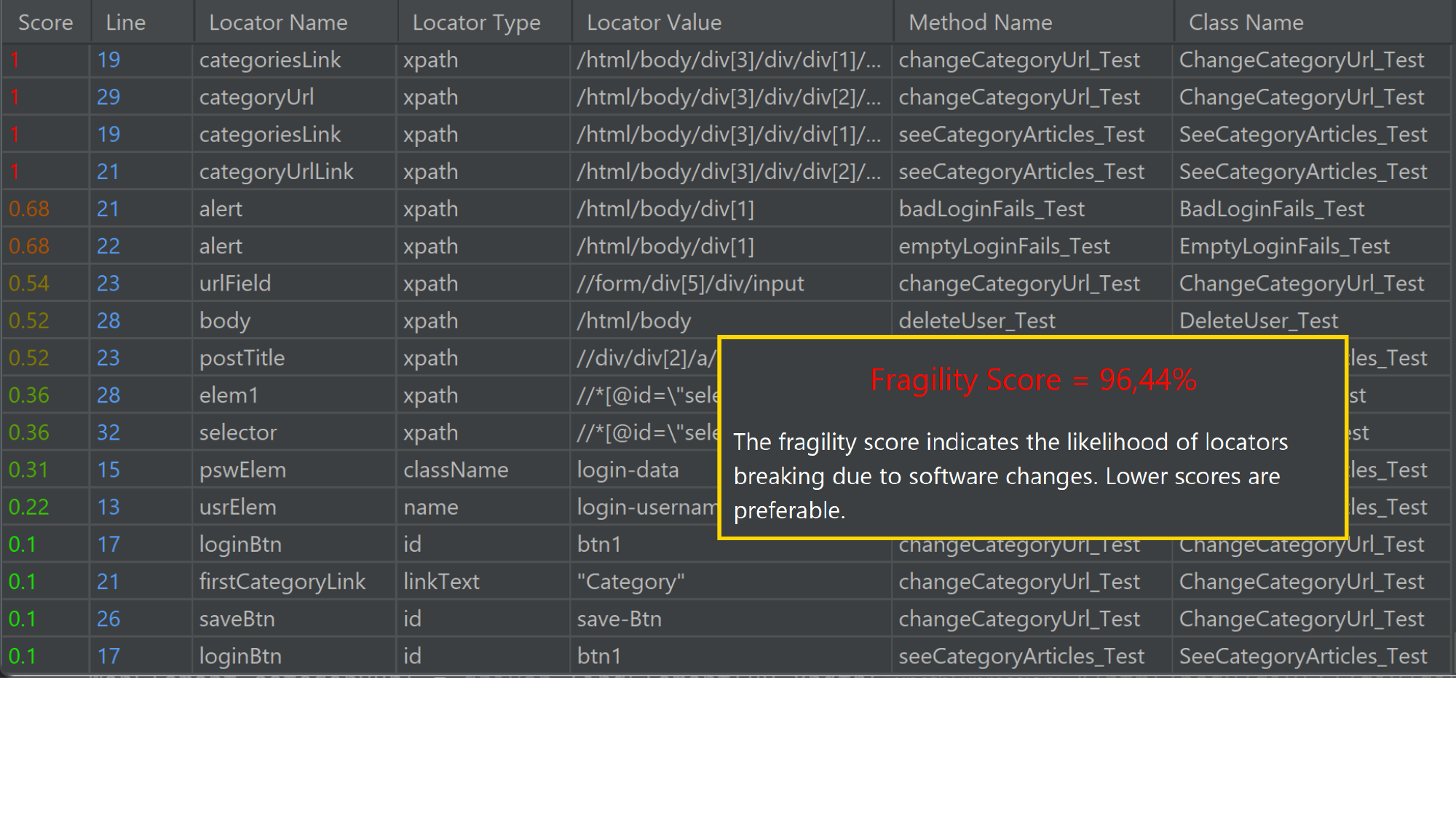}
    \vspace{-7mm}
    \caption{\plugin Fragility Window.}
    \label{fig:fragility-window}
    \vspace{-5mm}
\end{figure}

At the start of the plugin, and after each change on locators and Page Objects, the two main GUI windows of \plugin are displayed and refreshed, that are the \textit{Gamification} window (Figure \ref{fig:gamification-window}) and the \textit{Fragility} window (Figure \ref{fig:fragility-window}) about the estimation of locators fragility. The fragility of the locator is strictly based on the best practices listed in Table~\ref{tab:guidelines} and is a measure from 0 to 1 determining whether a locator is likely to break during software evolution. For example, a locator based on an absolute XPath value fails rule \textbf{L5}, resulting in a high fragility score and therefore a fragility index close to 1.  The algorithm that computes the fragility coefficient for a given locator assigns a different fragility coefficient to different types of locators (e.g. id, className etc.), and if the locator is an XPath it computes a fragility coefficient based on the types of predicates used in the XPath, and its length. Although it has several differences, the algorithm was inspired by the XPath generation algorithm used by Sidereal~\cite{leotta2021sidereal}.

The \textit{Gamification} window  is organized into three main panels: \textbf{User Info} panel (label 1 in Figure~\ref{fig:gamification-window}), \textbf{Dailies} panel (label 2), and \textbf{Achievements} panel (label 3). 
The \textbf{User Info} panel lists information about the user, such as name, propic, level, and current experience points. The \textbf{Dailies} panel lists the dailies assigned to the user for completion. The \textbf{Achievements} panel lists the unlocked and ongoing achievements. Finally, a daily assignment mode (label 4) at the bottom allows the selection of the daily assignment strategy, which can be \emph{RANDOM}, \emph{TARGETED} or \emph{INCLUSIVE}. 
In \emph{RANDOM} assignment mode, dailies are randomly assigned to the user and are expected to be completed within 24 hours, after which they expire and new ones are assigned. As randomness may result in infeasible dailies assigned, the user is offered to discard each daily once per day. In \emph{TARGETED} assignment mode, dailies are assigned to the user based on actual issues found in locators and Page Objects from analyzing the test suite according to the set of quality metrics listed in Table~\ref{tab:guidelines}. Since some issues might not be fixed, the user can flag as \textit{infeasible} any listed issue. In \emph{INCLUSIVE} assignment mode, dailies are assigned to the user based on  an automated DOM inspection of the Web application under test, that looks for potentially useful elements that have not been used in the tests yet. This mode is still under development.

The \textit{Fragility} window lists the locators found in the test suite, ordered by descending fragility score, providing information about the locator strategy implemented (e.g., type) and the position within the test suite. Further, an overall fragility score is shown and updated after each code change.

Once the main windows are shown, the user will be asked to complete the assigned tasks, by loading any previously produced progress that needs to be validated through test execution. 
Two listeners are registered to capture static and dynamic perspectives of the test suite, respectively, managing the changes affecting the code (e.g., re-evaluating the fragility scores of modified locators, or collecting the Page Objects failing the best practices of Table~\ref{tab:guidelines}) and the outcomes of test executions, eventually reactivating the whole process sketched in Figure~\ref{fig:testquest}.

The user will then make changes to the test artifacts and execute tests to validate these changes, resulting in task completion, periodically receiving new tasks to complete, and getting notified about any progress. The user will be able to interact with \plugin GUI to perform actions, such as editing profile data (e.g., change their propic), discarding dailies to get new ones, and viewing more details about the dailies to complete.
Notice that only positively exercised locators and Page Objects within executed tests are considered valid for progression. For instance, if a test case fails with a stacktrace indicating a line of failure in the test, any locator or Page Object method used by the test following that line will not be considered for the progression computation until the test passes.  
\plugin stores user data in XML files and periodically saves progress, including snapshots of locator and Page Object states, partially completed tasks, issues that have been fixed but require validation through test executions, and more.

%\DC{inserire esempio d'uso: daily assegnata con piccola spiegazione del problema (targeted) e mouse over, daily completata mostrando azione compiuta, reward ottenuta con progression su GUI e dettaglio su feasible e show/pending}

To give an example of practical use of \plugin, we consider a test suite that uses only absolute XPaths to locate the Web elements. The goal of the user will be to refactor the test suite with more robust XPaths. For this goal, when used in \emph{TARGETED} mode, \plugin proposes the daily "Convert the following absolute XPath locator to relative". When clicking the "Show" button (panel 2 in Figure \ref{fig:gamification-window}), all the absolute XPath locators in the test suites will be listed %in the following format (\texttt{variableName [ClassName.methodName, lineNumber]}) 
to be easily identifiable by the user. When the user changes an absolute XPath to a relative one (for example, from \texttt{\small /html/body/div[3]/div/div/form/div[1]/input} to \texttt{\small //*[@id="email"]}), and positively executes the containing test (i.e., no failures occur), the corresponding reward (e.g., experience points) will be assigned to the user, the changed locator will be removed from the list of locators proposed by the daily, and the overall fragility score will be updated in the Fragility Window (Figure~\ref{fig:fragility-window}).

For what concerns the usage of \plugin in adopting the Page Object model, a possible daily that may be proposed by \plugin is "Move from tests to Page Object methods the following locators". This daily is proposed when \plugin finds usages of locators in the bodies of the test scripts rather than in Page Object methods. To gain the corresponding reward, the user will have to move the listed locators in the Page Objects bodies, and then successfully run the containing test script.

%\section{Test Suites}\label{ch:testSuites}\input{40-testSuites}
%\section{Related Works}\label{sec:related}\input{60-related}
\section{Conclusion and Future Work}\label{sec:conclusion}%Web applications play a central role in our daily lives, requiring effective testing techniques to keep test artifacts aligned with rapid software changes. Although several automated testing techniques and frameworks have been developed to ease this process, locators - key elements used to identify widgets in Web pages - remain a leading cause of test failures. Moreover, the Page Object pattern, which promotes maintainability and abstraction, is often neglected or implemented without full adherence to established best practices, leading to time-consuming and repetitive fixes.
In this paper, we presented \plugin, a plugin for IntelliJ IDEA that applies gamification to Web testing, aiming to improve test quality by rewarding the use of robust locators and adherence to Page Object best practices.

As future work, we are planning to support and implement additional best practices targeting the test structure (e.g., a test case workflow must be short and simple) and conduct an experiment involving students and professionals to evaluate the effectiveness of \plugin in test quality and maintainability on real case studies. 

%\section*{Acknowledgments}
%This study was partially carried out within the ``EndGame - Improving End-to-End Testing of Web and Mobile Apps through Gamification'' project (2022PCCMLF) – Next Generation EU within the PRIN 2022 program (D.D.104 - 02/02/2022 Ministero dell’Università e della Ricerca). This manuscript reflects only the authors’ views and opinions and the Ministry cannot be considered responsible for them. 

%\bibliographystyle{abbrv}
\bibliographystyle{IEEEtran}
\def\IEEEbibitemsep{0.0pt plus .0pt}
\bibliography{99-refs}

\end{document}